\documentclass[aps,prl,twocolumn,groupedaddress]{revtex4}
\usepackage{graphicx}

\usepackage{natbib}
\begin{document}


\title{Spheromak formation and sustainment by tangential boundary flows}


\author{Pablo Luis Garc\'ia-Mart\'inez}
\author{Ricardo Farengo}
\affiliation{Instituto Balseiro and Centro At\'omico Bariloche}


\date{\today}

\begin{abstract}
The nonlinear, resistive, 3D magnetohydrodynamic equations are solved
numerically to demonstrate the possibility of forming and sustaining a
spheromak by forcing tangential flows at the plasma boundary.
The method can by explained in terms of helicity injection and differs
from other helicity injection methods employed in the past. Several
features which were also observed in previous dc helicity injection
experiments are identified and discussed.
\end{abstract}

\maketitle


Spheromak plasmas have been formed using several different techniques
\citep{jar94,jar05}.
The existence of multiple formation methods clearly shows that the
spheromak is a preferred (lowest energy) state toward which a
magnetohydrodynamic (MHD) system naturally evolves when the appropriate
boundary conditions are imposed.
The physical process responsible for the formation is magnetic
relaxation: on the time scale of MHD instabilities the plasma relaxes
to the minimum energy state compatible with  its magnetic helicity
content (which remains approximately constant) \cite{tay86}.
Once formed, the spheromak will decay in a resistive time scale
because resistivity does dissipate magnetic helicity.
For this reason, the sustainment of the configuration during times
longer than the resistive decay time requires some helicity injection
method.
Since relaxation operates on a shorter time scale, the spheromak
configuration is maintained regardless of the details of the
specific helicity injection mechanism.
Some particular examples are the coaxial helicity injection method
(CHI) \cite{jar83,bro92,hoo99}, the merging of helicity-carrying
filaments (MHF) \cite{woo03} and the helicity injected torus with
steady inductive helicity injection (HIT-SI) \cite{jar06}.

In this Letter we report the first evidence coming from nonlinear,
resistive, 3D MHD numerical simulations that demonstrate the
possibility of forming and sustaining a spheromak by forcing
tangential flows at the plasma boundary.
The method can by explained in terms of helicity injection and differs
from other helicity injection methods employed in the past (CHI, MHF
and HIT-SI).

An enhanced helicity injection mode was recently reported in spheromak
experiments with large plasma rotation \cite{wan07}.
Althought not analyzed in terms of boundary flows, this observation
could support the feasibility of the mechanism proposed and studied in
this Letter.

We model the plasma using the resistive MHD equations with finite
viscosity and zero $\beta$. 
The evolution equations for ${\bf u}$ and ${\bf B}$ are:
\begin{eqnarray}
  \label{eqmhd1}
  \partial_t {\bf u} + {\bf u} \cdot \nabla {\bf u} &=& 
  ({\bf J} \times {\bf B})/\rho_0 + \nu \nabla \cdot \Pi, \\
  \label{eqmhd2}
  \partial_t {\bf B} &+& \nabla \times {\bf E} = 0,
\end{eqnarray}
where ${\bf E}=-{\bf u \times B}+\eta{\bf J}$ and $\Pi=({\bf \nabla
 u+\nabla u}^T) - (2/3) \times (\nabla\cdot{\bf u})$ (see
 Ref. \cite{gar09b} for further details).
These equations are normalized with $a$ (chamber radius), $\psi_G$
 (imposed flux) and $c_A$ (Alfv\`en speed).
In addition, ${\bf B}$ and ${\bf J}$ are scaled with $\sqrt{\mu_0}$.
Time is expressed in units of the Alfv\`en time $\tau_A=a/c_A$.
The normalized resistivity $\eta$ and the kinematic viscosity $\nu$
 are set to $5\times 10^{-5}$.
With these values the resistive time is $\tau_r \sim 800$  (defined as
in Ref. \cite{izz03}).
The resulting system is solved with the Versatile Advection Code
\cite{tot96}.

The domain is a cylinder of elongation $h/a=1$, with perfectly
  conducting wall conditions (${\bf B \cdot n}=0$ and ${\bf J 
  \times n}=0$)
and vanishing velocity (${\bf u}=0$) at the cylindrical boundary
($r=a$) and the top end ($z=h$).
At the bottom end we impose the poloidal flux: $\psi(r,z=0) = C \psi_G
(r/a)^2 (1-r/a)^3$, where $C=28.935$ is a constant and $\psi_G$ is the
maximum flux imposed by the gun.
This geometry has been used to model the Spheromak Experiment (SPHEX)
\cite{bre02}.
If ${\bf u}=0$ is imposed at $z=0$ the full set of boundary conditions
leads to vanishing helicity and energy fluxes across the boundary.
Recently, these conditions have been applied to study the decay of
configurations representative of electrostatically driven spheromaks
with open field lines \cite{gar09c}.

Imposing tangential flows at a boundary intercepted by magnetic flux
may result in the injection of helicity, as can be inferred from the
equation $dH/dt = -2\int_V\eta{\bf J \cdot B}dV 
        - 2\oint_{\partial V} [({\bf A \cdot B})({\bf u \cdot n}) 
                             - ({\bf A \cdot u})({\bf B \cdot n})] dS$
(neglecting electrostatic fields).
The last term on the right gives the helicity injection produced by
motions of the footpoints of the penetrating (open) magnetic field
\cite{ber99}.
Boundary shearing of magnetic fields has been applied in the past to
study reconnection events in low-$\beta$ plasmas relevant to the solar
corona \cite{gal96}.

\begin{figure}
  \includegraphics[width=8.cm]{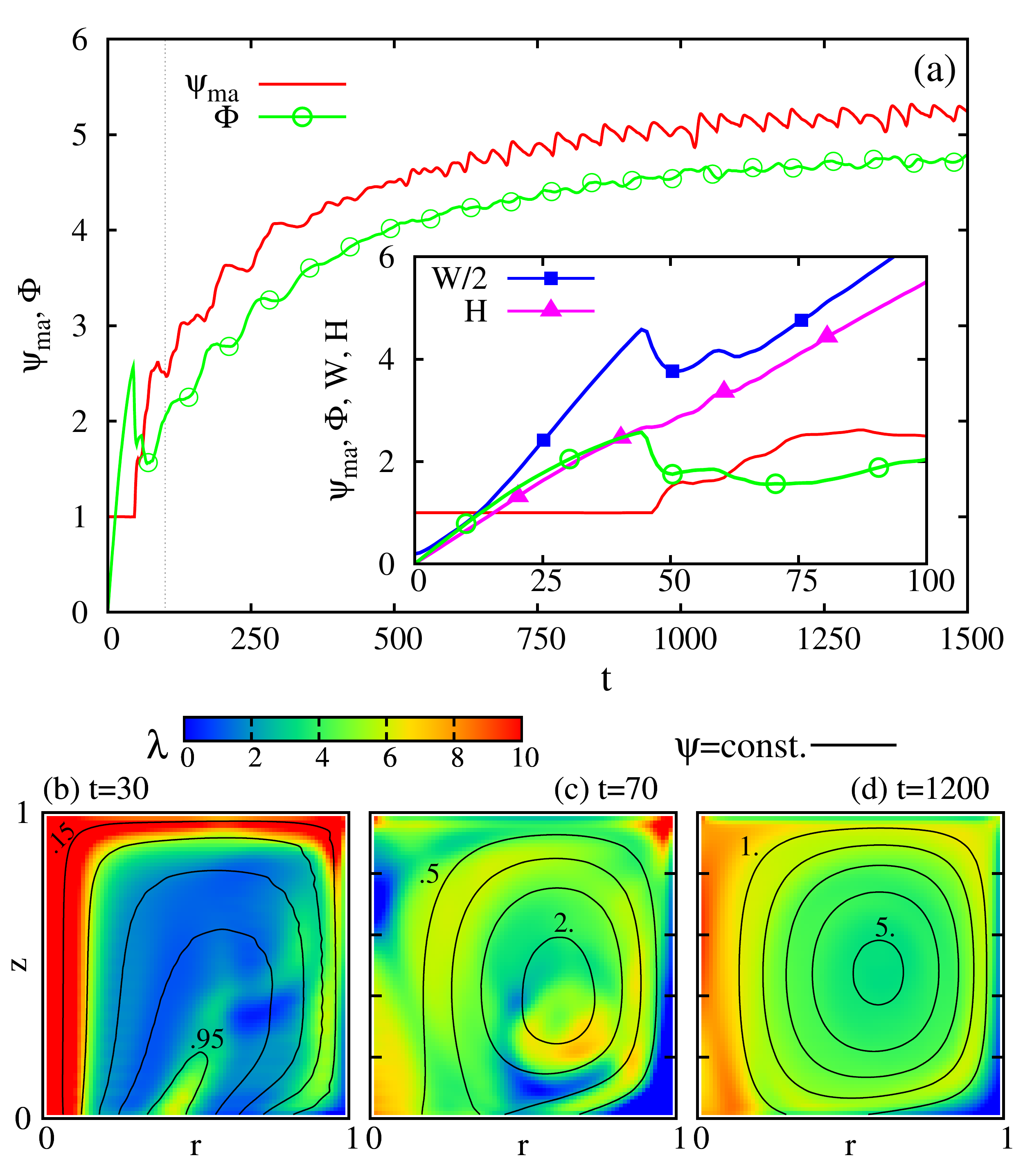}
  \caption{\label{form} (Color online) Evolution of $\psi_{ma}$ and
    $\Phi$ (a). In the inset, the evolution for $t<100$ is showed in
    more detail. In addition the (halved) magnetic energy and the
    helicity are shown. Colormaps of $\lambda$ and $\psi$ contours in
    the poloidal plane are shown at three different times (b)-(d).}
\end{figure}

\begin{figure}
  \includegraphics[width=8.5cm]{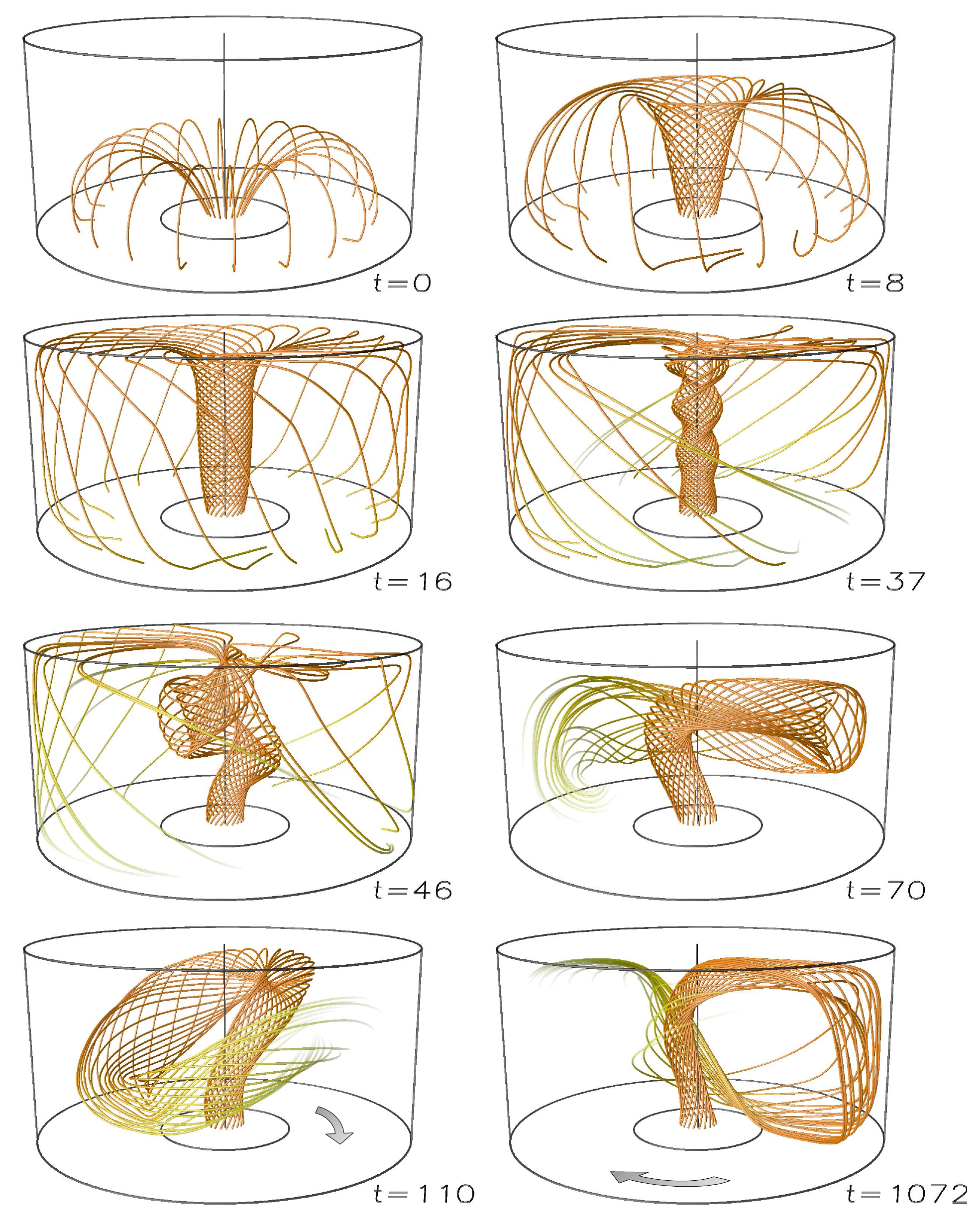}
  \caption{\label{colseq} (Color online) Twenty field lines followed
  from fixed positions at the bottom end, for eight different
  times. The opacity of the lines is gradually decreased when their
  length $L$ is longer than $4$, and they become completely
  transparent when $L \ge 5$. The structure observed at $t=110$ is
  almost in its final saturated state (observed at $t=1072$) and
  rotates clockwise (when viewed from top).}
\end{figure}

\begin{figure}
  \includegraphics[width=8.cm]{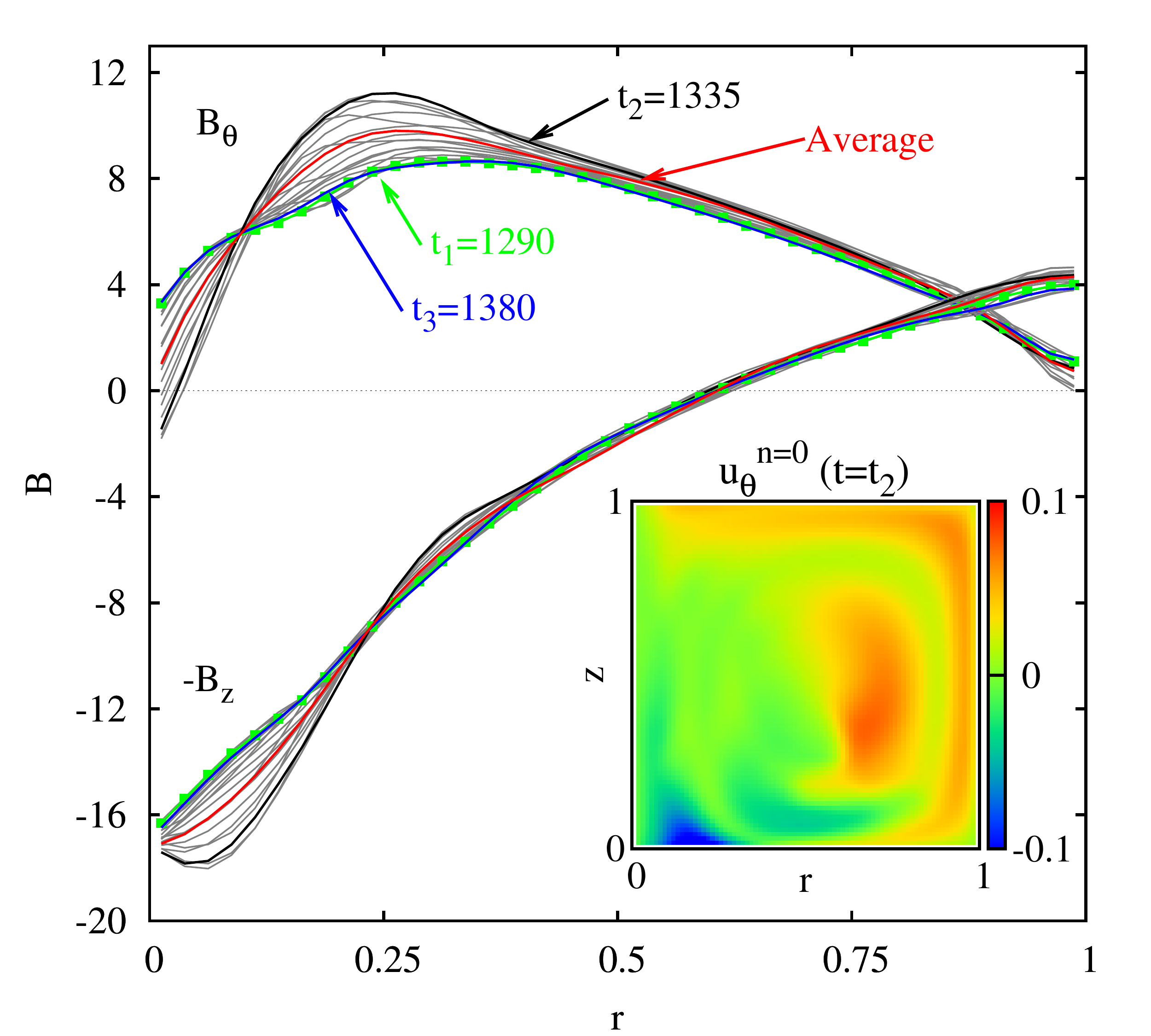}
  \caption{\label{brad} (Color online) Twenty instantaneous profiles
  (between $t_1$ and $t_3$) of $B_\theta$ and $B_z$, at $z=h/2$ and
    $\theta=0$. The profile of the time averaged fields is also
    shown. The $n=0$ component of the toroidal velocity at $t_2$ is
    shown in the inset.}
\end{figure}

\begin{figure}
  \includegraphics[width=8.5cm]{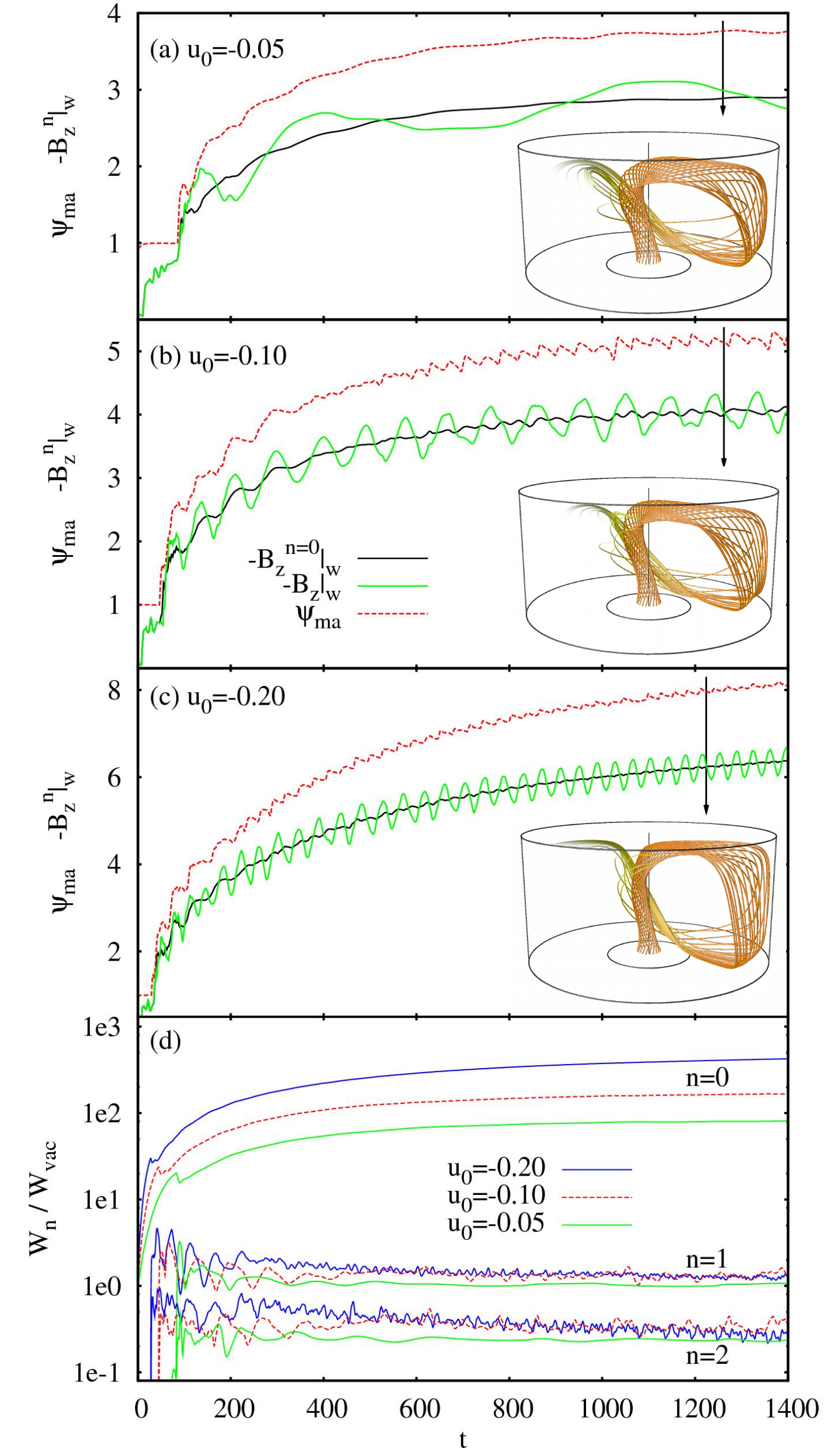}
  \caption{\label{fluc} (Color online) Evolution of $\psi_{ma}$ and
    the poloidal magnetic field near the wall (total $B_z|_w$ and
    $B_z^{n=0}|_w$) for three values of $u_0$ (a)-(c). Field lines
    showing the central open flux column are presented in each case at
    the time indicated by the arrow. Magnetic energy of the modes
    $n=0, 1$ and $2$ (d).}
\end{figure}

\begin{figure}
  \includegraphics[width=8.cm]{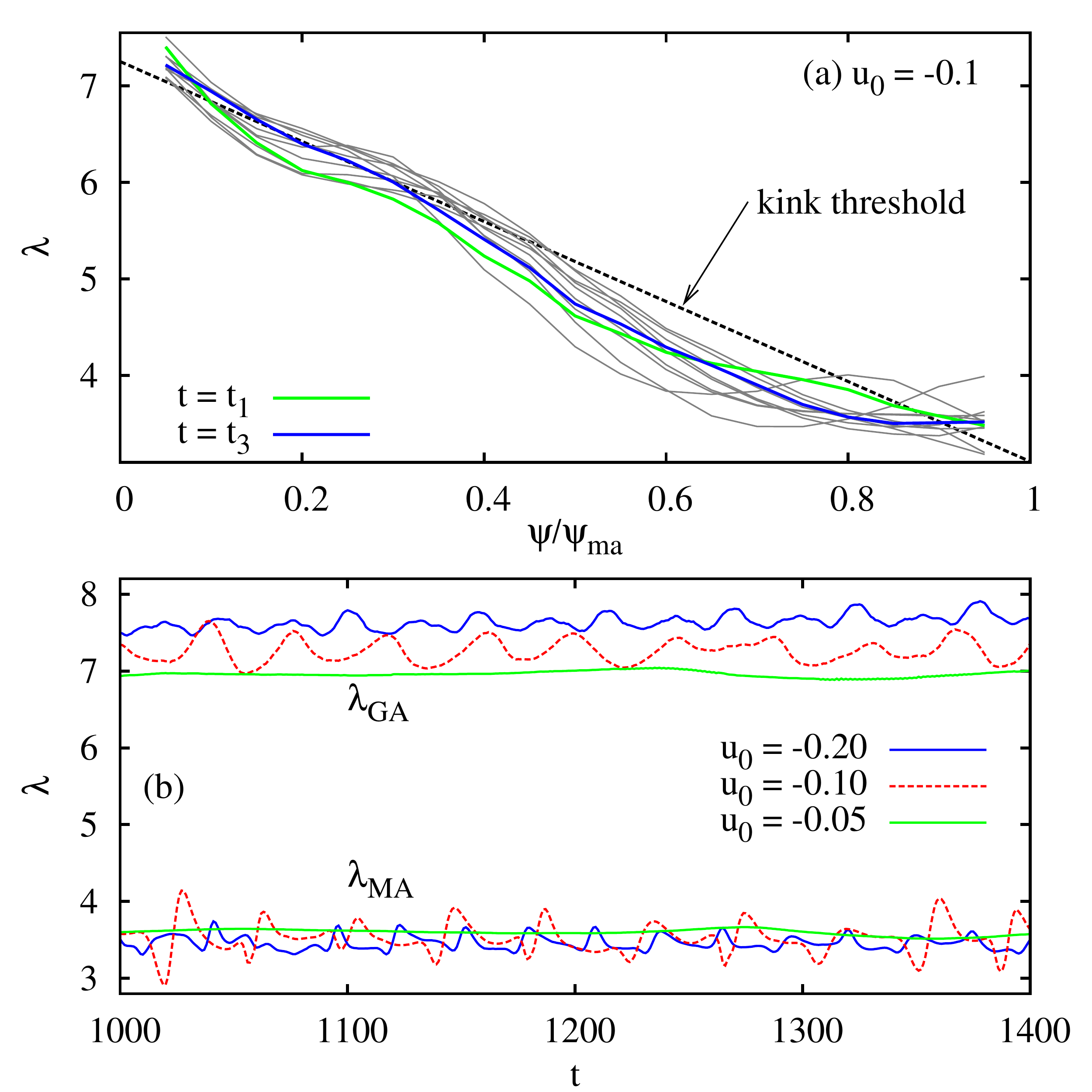}
  \caption{\label{lam} (Color online) $\lambda(\psi)$ profiles for the
    case with $u_0=-0.1$ at ten times between $t_1$ and $t_3$ (a). The
    kink instability threshold is indicated. Time evolution of 
    $\lambda$ at the geometric axis ($\lambda_{GA}$) and at the
    magnetic axis ($\lambda_{MA}$), for the three runs (b).}
\end{figure}

The computations presented in this Letter have $u_r=u_z=0$ and
$u_\theta = u_0$ max$[0,25(2/5-r)r]$, at $z=0$.
This flow produces helicity injection across the bottom end of the
cylinder because $A_\theta=\psi(r,z=0)/(2\pi r)$ there.
The initial condition is the vacuum magnetic field and zero
velocity inside the chamber.
Setting $u_0=-0.1$ we obtained the results shown in Figs. \ref{form},
\ref{colseq} and \ref{brad}.
The results of two more runs, with $u_0$ equal to $-0.05$ and $-0.2$,
are discussed in the context of Figs. \ref{fluc} and \ref{lam}.

The evolutions of the maximum poloidal flux ($\psi_{ma}$) and the
toroidal flux across the entire poloidal plane ($\Phi$) are shown in
Fig. \ref{form} (a).
For $t < 45$, $\psi_{ma}$ remains at the constant imposed value, while
there is a build up of $\Phi$.
Magnetic energy and helicity (computed using a gauge invariant
definition \cite{fin85}) are shown in the inset of Fig. \ref{form}
(a).
At $t \sim 45$, a relaxation event which produces significant energy
dissipation at approximately constant helicity and flux conversion
from toroidal to poloidal is clearly observed.
The kinetic energy $K$ (not shown) is small during the whole evolution 
($K/W$ has a peak of $0.1$ at $t \sim 50$ and remains below $0.01$
during sustainment).
Panels (b), (c) and (d) of Fig. \ref{form} show contours of $\psi$ and
colormaps of $\lambda$ (where $\lambda={\bf J \cdot B}/B^2$, is
computed using the $n=0$ component of the toroidal Fourier
decomposition of ${\bf B}$) at different times.
At $t=30$, $\lambda$ is strongly peaked near the geometric axis
and no closed $\psi$ contours are identified.
After the relaxation event ($t=70$) $\lambda$ is redistributed and
closed $\psi$ contours appear, indicating the formation of a spheromak
configuration.
The poloidal and toroidal fluxes continue to increase until $t \sim
700$ and afterwards a quasi stationary state is sustained for one
resistive time, indicating that a balance between helicity injection
and dissipation has been reached.
The $\psi$ and $\lambda$ distributions shown in Fig. \ref{form} (d)
are representative of this quasi-steady state.

It is well known that spheromak formation and sustainment necessarily
involves non-axisymmetric activity.
The magnetic field lines (followed from fixed positions at the
bottom end) plotted in Fig. \ref{colseq} clearly show that our
computations reproduce this feature.
The initial field lines of the vacuum solution ($t=0$) expand and
twist ($t=8$) forming a central current-carrying column ($t=16$).
As the current through the central column increases and the field
lines increase their twisting, the configuration eventually becomes
unstable ($t=37$) and the typical helical structure of a kink
instability quickly develops ($t=46$).
This instability, with dominant toroidal wavenumber $n=1$, saturates
at a relatively small and fixed amplitude ($W_{n=1}/W_{n=0} \sim 0.01$
during sustainment, see Fig. \ref{fluc}) causing the central column to
adopt an almost fixed helical shape which rotates as indicated in
Fig. \ref{colseq} ($t=110$ and $t=1072$).
Our results reproduce the experimental observation of a fixed rotating
helical structure with a strongly asymmetric return column 
\cite{duc97}.
In contrast with our case (driven purely by boundary flows), the
plasma in those experiments was driven electrostatically.
We note, however, that the presence of an inductive component in the
electric field when magnetic fluctuations were active, was also
reported \cite{duc97}.

It is important to note that the motion of this structure is produced
by the coherent oscillation of the $n=1$ mode and not by a rigid
rotation of the plasma.
This is shown in Fig. \ref{brad} where the radial profiles of
$B_\theta$ and $B_z$, at $z=h/2$ and $\theta=0$, are plotted for
several times between $t_1=1290$ and $t_3=1380$.
The profiles at $t_1$ and $t_3$ are almost identical, indicating
that the structure makes one turn every $90$ Alfv\`en times.
Such rotation is much slower than that expected from the imposed
$u_0=-0.1$ at $r=0.2$.
Furthermore, the plasma toroidal velocity reverses its sign as can be
observed in the inset of Fig. \ref{brad} (this velocity colomap is
plotted at $t_2=1335$ but it is representative of the flow pattern
during sustainment, $700<t<1500$).
This velocity reversal is also observed in the other two runs
presented in this work and it is evidently a feature of the saturated
state of the instability.
However, the specific reason for this velocity reversal is still not
yet understood.

The evolution of $\psi_{ma}$ and the poloidal magnetic field near the
wall (total $B_z|_w$ and $B_z^{n=0}|_w$) at $z=h/2$ and $\theta=0$ are
shown for three values of $u_0$ in Fig. \ref{fluc} (a)-(c).
The magnetic energy (relative to the initial energy) of the modes
$n=0$, $n=1$ and $n=2$ is shown in Fig. \ref{fluc} (d).
These results indicate that increasing the helicity injection rate
(which is linear in $u_0$) leads to configurations with higher flux
amplification and more energy in the $n=0$ mode.
Nevertheless, the energy associated to the $n>0$ modes (in particular
the $n=1$) saturates at a fixed amplitude, which is roughly
independent of $u_0$.
This implies a lower level of fluctuations relative to the $n=0$ mode
at higher helicity injection rates.
A similar saturation mechanism has also been observed in experiments
\cite{duc97,wil99}.

The oscillation of the fluctuation in the poloidal magnetic field near
the wall ($B_z|_w$) is clearly observed in Figs. \ref{fluc} (a)-(c).
As discussed previously, this corresponds to a coherent oscillation
which produces the rotation of the magnetic structures shown in the
insets of Figs. \ref{fluc} (a)-(c).
Note that the rotation frequency increases rapidly with the helicity
injection rate.
Another important observation is that the rotation frequency is not
the same, in general, that the frequency of the oscillating dynamo
term that amplifies $\psi_{ma}$ (which depends on the correlation of
the fluctuations of ${\bf u}$ and ${\bf B}$).
This is specially clear in Fig. \ref{fluc} (b).

The MHD activity maintains the configurations obtained for the three
different values of $u_0$ close to the kink stability boundary.
This is shown in Fig. \ref{lam}.
The $\lambda(\psi)$ profiles of the $n=0$ mode are plotted in
Fig. \ref{lam} (a), for the case with $u_0=-0.1$, at ten times between
$t_1$ and $t_3$ ($\psi$ is normalized with $\psi_{ma}$).
The indicated kink instability threshold is a linear $\lambda(\psi)$
profile, which has a slope $\alpha=-0.4$ \cite{kno86}.
Fig. \ref{lam} (b) shows the evolution of the $\lambda$ values at the
geometric axis ($\lambda_{GA}$) and at the magnetic axis
($\lambda_{MA}$) for the three runs.
The behavior observed here is similar to that described in previous
experiments \cite{kno86,wil99}.
The configuration evolves around the kink stability boundary as a
result of the competition between the external forcing, which tends to
steepen the $\lambda$ profile, and the relaxation, which tends to
flatten it.

In summary, we demonstrated the possibility of forming and sustaining
a spheromak by imposing tangential flows at the boundary.
Several aspects of the process were described.
The relationship between our results and the existing experimental
evidence was discussed.
The fact that our simulations reproduce many features observed in
electrostatic CHI experiments is interpreted as follows.
Since the spheromak is formed and sustained by the relaxation of an
unstable configuration, the dynamics of the process should be
independent of the details on how this configuration is driven
unstable.
In this sense, this Letter not only provides evidence on a new
spheromak formation and sustainment mechanism, but it also provides
valuable information pertaining to the dynamics of the kink
instability in electrostatically driven spheromaks.
To conclude, we note that althought our results were obtained for a
spheromak, boundary plasma flows could also be used to inject helicity
in other configurations (i.e. spherical tokamaks).

\begin{acknowledgments}
Financial support from the UNCuyo and the ANPCyT is
acknowledged.
P.L.G.-M. is supported by CONICET.
\end{acknowledgments}


\end{document}